# Mapping hybrid functional-structural connectivity traits in the human connectome

Enrico Amico [1,2] and Joaquín Goñi [1,2,3,*]

[1] School of Industrial Engineering, Purdue University, West-Lafayette, IN, USA
[2] Purdue Institute for Integrative Neuroscience, Purdue University, West-Lafayette, IN, USA
[3] Weldon School of Biomedical Engineering, Purdue University, West-Lafayette, IN, USA

[*] Correspondence: jgonicor@purdue.edu

## Abstract

One of the crucial questions in neuroscience is how a rich functional repertoire of brain states relates to its underlying structural organization. How to study the associations between these structural and functional layers is an open problem that involves novel conceptual ways of tackling this question. We here propose an extension of the Connectivity Independent Component Analysis (connICA) framework, to identify joint structural-functional connectivity traits.

Here, we extend connICA to integrate structural and functional connectomes by merging them into common "hybrid" connectivity patterns that represent the connectivity fingerprint of a subject. We test this extended approach on the 100 unrelated subjects from the Human Connectome Project. The method is able to extract main independent structural-functional connectivity patterns from the entire cohort that are sensitive to the realization of different tasks.

The hybrid connICA extracted two main task-sensitive hybrid traits. The first, encompassing the within and between connections of dorsal attentional and visual areas, as well as fronto-parietal circuits. The second, mainly encompassing the connectivity between visual, attentional, DMN and subcortical networks. Overall, these findings confirms the potential of the hybrid connICA for the compression of structural/functional connectomes into integrated patterns from a set of individual brain networks.

## Introduction

Brain Connectomics is a rapidly growing area of research (Bullmore & Sporns, 2009; Fornito, Zalesky, & Bullmore, 2016). It is based on the investigation of functional and structural connections in the human brain, modeled as networks (Bullmore & Sporns, 2009; Fornito et al., 2016; Sporns, 2011). In large-scale brain network models, nodes correspond to gray-matter regions (based on brain atlases or parcellations) while links or edges correspond to connections between the nodes. Structural connections are modeled from diffusion weighted imaging (DWI) data, by inferring the main white matter axonal pathways between brain region pairs through tractography algorithms (Fornito et al., 2016), normally denominated structural connectome or structural connectivity (SC) (Sporns, 2011). Functional connections are modeled from functional magnetic resonance imaging (fMRI) data, by measuring temporal statistical dependences between the estimated neural activity of brain region pairs while subjects are either at rest or performing a task in the scanner, usually defined as functional connectivity or functional connectome (FC) (Fox & Raichle, 2007; Friston, 2011)

The exponential growing of publicly available neuroimaging datasets in the last years has allowed researchers to make inferences on the different organization of brain networks in clinical and healthy populations, and to identify changes in these cohorts (Fornito, Zalesky, & Breakspear, 2015; Fornito et al., 2016), both at the structural and functional level. During the past years, many efforts have also been made to address one of the crucial questions in Brain Connectomics. That is, how a rich functional repertoire of brain states relates to its underlying structural organization, especially at the large scale of cortical/sub-cortical grey matter modules and white matter fiber-bundles (Falcon, Jirsa, & Solodkin, 2016; Goñi et al., 2014; C. J. Honey et al., 2009; Christopher J. Honey, Kötter, Breakspear, & Sporns, 2007; Christopher J. Honey, Thivierge, & Sporns, 2010).

The study of the associations between these structural and functional layers (Mišić et al., 2016) is difficult to accomplish due to several factors. One factor is related to obtaining individual accurate connectivity patterns. This involves: the design of magnetic resonance imaging (MRI) sequences for structural imaging, diffusion weighted imaging, and functional MRI (fMRI); the development of processing pipelines to process MRI data; a brain parcellation or atlas to reduce the dimensionality from gray matter voxels to brain regions and criteria to estimate levels of structural and functional coupling.

Another aspect relates to the inter-subject variability of these two modalities. The identification of group-level structure-function relationships (Mišić et al., 2016) may become an even more powerful approach if individual estimations were taken into account. As a matter of fact, it has recently been shown that the "individual fingerprint" of a functional connectome (Finn et al., 2015) is a key property for investigating further inferences and links between connectomics and genetic, demographic or clinical variables (Shen et al., 2017). The recent trend goes therefore towards working at the single subject level, and towards the refinement and improvement of this individual signature in a individual human connectome (Amico & Goñi, 2017). In this sense, providing not only group-level SC or co-varying SC/FC patterns but also their individual estimations is an important step forward. Lastly, the vast amount of information contained in both functional and structural connectomes is problematic for the investigation of joint FC and SC patterns. In this scenario, the researcher has to extract and compress informative features from hundreds of functional and structural connectomes separately, from either healthy or clinical populations, and then come up with creative ways to merge the extracted functional information with its structural counterpart, or find ways to compress them in some integrative framework.

We here define an extension of our recently proposed Connectivity-based Independent Component Analysis (i.e. connICA, (Amico et al., 2017)) technique, to overcome the aforementioned issues. The connICA methodology implements Independent Component Analysis (ICA) for the extraction of robust independent functional connectivity patterns from a set of individual functional connectomes, without imposing any a priori data stratification into groups (Amico et al., 2017). Here, we extend connICA to include both structural and functional connectomes, by merging them into a common "hybrid" matrix (see scheme at Fig. 1) that includes both the structural and functional fingerprint of each subject. We test this extended approach on the 100 unrelated subjects taken from the Human Connectome Project (details on the project available at http://www.humanconnectome.org/) and evaluated it for two brain parcellations We here show how this method is able to extract main independent structure-function couplings with individual estimations for the entire population of subjects, and to disentangle the joint functional-structural sub-systems that are sensitive to different functional tasks (including also resting state).

These findings confirm the potential of the hybrid connICA for the compression of meaningful information out of a set of heterogeneous brain networks based on both functional and structural connectomes while capturing individual differences. We conclude by discussing limitations and potential future directions for this methodology.

## Materials and Methods

**Dataset.** The functional and structural dataset used in this work is from the Human Connectome Project (HCP, http://www.humanconnectome.org/), Release Q3. Per HCP protocol, all subjects gave written informed consent to the Human Connectome Project consortium. Below is the full description of the acquisition protocol and processing steps.

**HCP: functional data.** We used fMRI runs from the 100 unrelated subjects of the HCP 900 subjects data release (D. C. Van Essen et al., 2012; David C. Van Essen et al., 2013). The fMRI resting-state runs (HCP filenames: rfMRI_REST1 and rfMRI_REST2) were acquired in separate sessions on two different days, with two different acquisitions (left to right or LR and right to left or RL) per day (Glasser et al., 2013; D. C. Van Essen et al., 2012; David C. Van Essen et al., 2013). The seven fMRI tasks were the following: gambling (tfMRI_GAMBLING), relational (tfMRI_RELATIONAL), social (tfMRI_SOCIAL), working memory (tfMRI_WM), motor (tfMRI_MOTOR), language (tfMRI_LANGUAGE, including both a story-listening and arithmetic task) and emotion (tfMRI_EMOTION). The working memory, gambling and motor task were acquired on the first day, and the other tasks were acquired on the second day (Barch et al., 2013; Glasser et al., 2013). The HCP scanning protocol was approved by the local Institutional Review Board at Washington University in St. Louis. For all sessions, data from both the left-right (LR) and right-left (RL) phase-encoding runs were used to calculate connectivity matrices. Full details on the HCP dataset have been published previously (Barch et al., 2013; Glasser et al., 2013; S. M. Smith et al., 2013).

**HCP: structural data.** We used DWI data from the same 100 unrelated subjects of the HCP 900 subjects data release (D. C. Van Essen et al., 2012; David C. Van Essen et al., 2013).

The diffusion acquisition protocol is covered in detail elsewhere (Glasser et al., 2013; Sotiropoulos et al., 2013; Uğurbil et al., 2013). Below we mention the main characteristics. Very high-resolution acquisitions (1.25 mm isotropic) were obtained by using a Stejskal–Tanner (monopolar) (Stejskal & Tanner, 1965) diffusion-encoding scheme. Sampling in q-space was performed by including 3 shells at b=1000, 2000 and 3000 s/mm2. For each shell corresponding to 90 diffusion gradient directions and 5 b=0's acquired twice were obtained, with the phase encoding direction reversed for each pair (i.e. LR and RL pairs). Directions were optimized within and across shells (i.e. staggered) to maximize angular coverage using the approach of (Caruyer et al., 2011)(http://www-sop.inria.fr/members/Emmanuel.Caruyer/q-space-sampling.php), and form a total of 270 non-collinear directions for each PE direction. Correction for EPI and eddy-current-induced distortions in the diffusion data was based on manipulation of the acquisitions so that a given distortion manifests itself differently in different images (Andersson, Skare, & Ashburner, 2003). To ensure better correspondence between the phase-encoding reversed pairs, the whole set of diffusion-weighted (DW) volumes is acquired in six separate series. These series were grouped into three pairs, and within each pair the two series contained the same DW directions but with reversed phase-encoding (i.e. a series of Mi DW volumes with RL phase-encoding is followed by a series of Mi volumes with LR phase-encoding, i = [1,2,3]).

**Brain atlases.** We employed a cortical parcellation into 360 brain regions as recently proposed by Glasser et al. (Glasser et al., 2016). For completeness, 14 sub-cortical regions were added, as provided by the HCP release (filename "Atlas_ROI2.nii.gz"). To do so, this file was converted from NIFTI to CIFTI format by using the HCP workbench software (Glasser et al., 2013; Marcus et al., 2011)(command *-cifti-create-label* http://www.humanconnectome.org/software/connectome-workbench.html). An additional parcellation scheme was also evaluated (Destrieux, 164 brain regions (Destrieux, Fischl, Dale, & Halgren, 2010; Fischl et al., 2004), as available in FreeSurfer).

**HCP preprocessing: functional data.** The HCP functional preprocessing pipeline (Glasser et al., 2013; S. M. Smith et al., 2013) was used for the employed dataset. This pipeline included artefact removal, motion correction and registration to standard space. Full details on the pipeline can be found in (Glasser et al., 2013; S. M. Smith et al., 2013). The main steps were: spatial ("minimal") pre-processing, in both volumetric and grayordinate forms (i.e., where brain locations are stored as surface vertices (S. M. Smith et al., 2013)); weak highpass temporal filtering (> 2000s full width at half maximum) applied to both forms, achieving slow drift removal. MELODIC ICA (Jenkinson, Beckmann, Behrens, Woolrich, & Smith, 2012) applied to volumetric data; artifact components identified using FIX (Salimi-Khorshidi et al., 2014). Artifacts and motion-related time courses were regressed out (i.e. the 6 rigid-body parameter time-series, their backwards-looking temporal derivatives, plus all 12 resulting regressors squared) of both volumetric and grayordinate data (S. M. Smith et al., 2013).

For the resting-state fMRI data, we also added the following steps: global gray matter signal was regressed out of the voxel time courses (Power et al., 2014); a bandpass first-order Butterworth filter in forward and reverse directions [0.001 Hz, 0.08 Hz] (Power et al., 2014) was applied (Matlab functions *butter* and *filtfilt*); the voxel time courses were z-scored and then averaged per brain region, excluding outlier time points outside of 3 standard deviation from the mean, using the workbench software (Marcus et al., 2011) ( workbench command *-cifti-parcellate* ). For task fMRI data, we applied the same above mentioned steps, with a less restrictive range for the bandpass filter [0.001 Hz, 0.25 Hz].

Pearson correlation coefficients between pairs of nodal time courses were calculated (MATLAB command *corr*), resulting in a symmetric connectivity matrix for each fMRI session of each subject. Functional connectivity matrices from the left-right (LR) and right-left (RL) phase-encoding runs were averaged to improve signal-to-noise ratio. The functional connectomes were kept in its signed weighted form, hence neither thresholded nor binarized. This was done for the two parcellations described above, namely Glasser with subcortical regions added (giving a total of 374 brain regions) and Destrieux (164 brain regions).

Finally, the resulting individual functional connectivity matrices were ordered (rows and columns) according to 7 resting-state cortical sub-networks (RSNs) as proposed by Yeo and colleagues (Yeo et al., 2011). For completeness, an 8th sub-network including the 14 HCP sub-cortical regions was added (as analogously done in recent paper (Amico et al., 2017)).

**HCP preprocessing: structural data.** The HCP DWI data were processed following the MRtrix3 (Tournier, Calamante, & Connelly, 2012) guidelines (http://mrtrix.readthedocs.io/en/latest/tutorials/hcp_connectome.html). In summary, we first generated a tissue-segmented image appropriate for anatomically constrained tractography (ACT (R. E. Smith, Tournier, Calamante, & Connelly, 2012), MRtrix command *5ttgen*); we then estimated the multi-shell multi-tissue response function ((Christiaens et al., 2015), MRtrix command *dwi2response msmt_5tt*) and performed the multi-shell, multi-tissue constrained spherical deconvolution ((Jeurissen, Tournier, Dhollander, Connelly, & Sijbers, 2014), MRtrix *dwi2fod msmt_csd*); afterwards, we generated the initial tractogram (MRtrix command *tckgen*, 10 million streamilines, maximum tract length = 250, FA cutoff = 0.06) and applied the successor of Spherical-deconvolution Informed Filtering of Tractograms (SIFT2, (R. E. Smith, Tournier, Calamante, & Connelly, 2015)) methodology (MRtrix command *tcksift2*). Both SIFT

(R. E. Smith, Tournier, Calamante, & Connelly, 2013) and SIFT2 (R. E. Smith et al., 2015, p. 2) methods provides more biologically meaningful estimates of structural connection density. SIFT2 allows for a more logically direct and computationally efficient solution to the streamlines connectivity quantification problem: by determining an appropriate cross-sectional area multiplier for each streamline rather than removing streamlines altogether, biologically accurate measures of fibre connectivity are obtained whilst making use of the complete streamlines reconstruction (R. E. Smith et al., 2015). Then, we mapped the SIFT2 outputted streamlines onto the 374 chosen brain regions (360 from Glasser et al. (Glasser et al., 2016) brain atlas plus 14 subcortical regions, see Brain Atlases section) to produce a structural connectome (MRtrix command *tck2connectome*). Finally, a $\log_{10}$ transformation (Fornito et al., 2016) was applied on the structural connectomes to better account for differences at different magnitudes. Consequently, SC values ranged between 0 and 5 on this dataset. In order to test the method with a different parcellation scheme, we performed the same mapping steps from the streamlines to a second parcellation (Destrieux, 164 brain regions (Destrieux et al., 2010; Fischl et al., 2004), as available in FreeSurfer).

**Hybrid connICA: independent component analysis of joint functional and structural connectomes**

The potential of multidimensional feature extraction from different neuroimaging modalities has been already introduced and explored (V. D. Calhoun et al., 2006; Vince D. Calhoun et al., 2006; Vince D. Calhoun, Liu, & Adalı, 2009) in the fMRI domain. Recently, applications of independent component analysis (ICA, (Hyvärinen & Oja, 2000)) in the functional connectome domain (Amico et al., 2017; Kessler, Angstadt, & Sripada, 2016) and in joint patterns of functional connectomes and grey/white matter volumes (Kessler, Angstadt, Welsh, & Sripada, 2014) have been investigated. Here we propose a framework that allows for the extraction of joint connectivity traits from a set of functional and structural connectomes, based on the extension of our recently proposed Connectivity-based Independent Component Analysis (connICA, (Amico et al., 2017)), here named "hybrid connICA". Below is the detailed description of the hybrid connICA scheme.

The first step relates to uniforming the different distributions of FC values (Pearson's correlation values ranging between -1 to 1) and SC values (after $\log_{10}$ transformation, ranging between 0 and 5). There are several options to normalize FC and SC connections in the same range: here we proceeded as follows. For each pair of nodes *i* and *j* directly connected in the SC matrix, we evaluated their "structural correlation". That is, the Pearson's correlation coefficient between the $i_{th}$ and $j_{th}$ row of the structural connectome. Iterating this procedure over all connected pairs gives a correlation matrix of a structural connectome (see Fig. S2 of the supplementary material). The values in this matrix range between [-1 1], with negative values indicating two nodes that are connected antagonistically to the rest of the network, and positive indicating high similarity in their structural connections with the rest of the brain network. This solution, similarly to matching index (Rubinov & Sporns, 2010), provides several advantages: does not change the general properties of the SC (Fig. S2) and it also allows to have functional and structural connectomes in the same range ([- 1 1]). However, this transformation also changes the SC matrix structure from sparse to full. Therefore, in order for this correlation matrix to be representative of the real structural architecture of a human brain, we only considered the correlation values corresponding to structurally connected pairs of brain regions (that is, edges with non-zero values in all the SCs of the population; this corresponds to approximately 21% of all possible pairs, see Fig. S2).

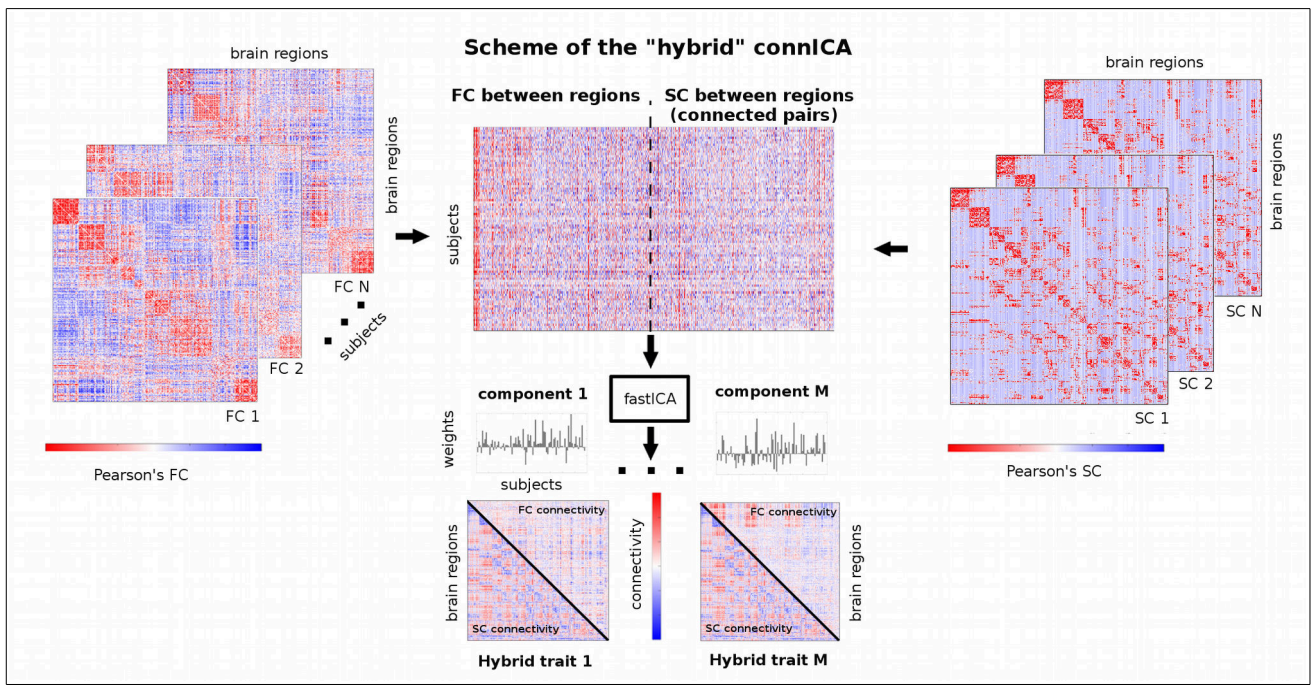

**Fig.1: Workflow scheme of the hybrid connICA.** The upper triangular of each individual functional connectivity (FC) matrix (left) and lower triangular of each correspondent structural connectivity profile (SC) are added to a matrix where rows are the subjects and columns are their vectorized hybrid (structural-functional) connectivity patterns. Note that for SC, only connected pairs across all subjects were included (see Methods for details). The ICA algorithm extracts the M independent components (i.e. hybrid traits) associated to the whole population and their relative weights across subjects. Colorbars indicate positive (red) and negative (blue) connectivity values: Pearson's correlation coefficient values in the case of individual FC and SC matrices (left and right side of scheme), and unitless connectivity weights in the case of hybrid traits (bottom of the scheme).

The second step is similar to the standard connICA approach (Amico et al., 2017): the input of the hybrid connICA consists of all the individual FC and SC profiles embedded into a "hybrid" dataset matrix where each row contains all the edges of the upper triangular part of an individual FC matrix (first half) and the correspondent lower triangular part of the SC matrix from the same individual (second half, see scheme at Fig. 1). Note that, due to symmetry of Pearson's correlations on FC and SC, taking the upper or the lower part of both matrices is just conventional. In this cross-sectional study, we selected 10 different subjects per task (7 tasks and 1 resting-state, see HCP: functional data section), ending up with a hybrid matrix of 80 hybrid connectivity profiles. Each profile represents the unique hybrid connectivity signature (both structural and functional) of a human brain (Fig. 1). Note that this method is insensitive to the ordering of the columns on the input hybrid matrix (i.e. it does not affect the results obtained).

Before running the ICA algorithm, dimensionality reduction on the dataset was obtained by applying principal component analysis (PCA, (Jolliffe, 2014)) on the hybrid matrix. The advantage of applying PCA before ICA for noise filtering and dimensionality reduction in order to avoid over-fitting has already been shown, both by the machine learning (Särelä & Vigário, 2003) and neuroimaging communities (V. D. Calhoun et al., 2006; Kessler et al., 2014). Recently, we also showed that PCA decomposition and subsequent reconstruction of functional connectomes can increase individual identifiability in a population, by retaining an optimal number of principal components (which usually explained the 90% of the variance in the functional data employed, see (Amico & Goñi, 2017)). Here, we applied PCA to compress and reduce the dimensionality of the data, by keeping the principal components explaining 90% of the variance of the initial hybrid data. Indeed, since the hybrid input matrix is highly redundant (due to high similarity in structural healthy connectomes, as well as task-based FCs), 40 components explained 90% of variance in the data (see Fig. S1 of the supplementary material).

Next, ICA decomposition of the PCA-reconstructed hybrid matrix was applied by running the FastICA algorithm (Hyvarinen, 1999). Similarly to connICA (Amico et al., 2017), the output of the hybrid connICA consists of two vectors per component. The first output vector will be referred to as hybrid-trait, which represents an independent pattern of joint functional-structural connectivity, common to the whole population. The second output vector is the weight of the hybrid-trait on each subject, which quantifies the prominence or presence of the trait in each individual connectivity hybrid profile (both functional and structural). This methodology allows for compressing the information contained in a population of structural and functional connectomes into few connectivity traits and unique individual weights associated to them. This can greatly ease the process of making inferences between the hybrid-connectivity subsystems present in a single-subject structural-functional connectivity profile and genetic, demographic or clinical variables at hand.

Given the non-deterministic nature of the FastICA decomposition into components (Hyvarinen, 1999), it is very important to run it several times and only select the most robust outcomes, in this case hybrid traits. We evaluated the robustness of the traits over 100 FastICA runs, as in (Amico et al., 2017). A bootstrap technique was used to accurately estimate the hybrid traits from the 100 subjects pool of the HCP dataset (see HCP data section for details). At every run, random samples comprising hybrid profiles from 80 different subjects (10 subjects per task and resting-state) were performed. This was meant to avoid results driven by a small subset of the population. Finally, a hybrid-trait was considered robust when it appeared (correlation of 0.5 or higher across runs) in at least 50% of the runs and its representation consisted of the average across all its appearances over the 100 runs

The last point worth mentioning of the procedure relates to the number of independent component chosen. There is not a gold standard for this choice: it usually depends on heuristic measures and the dataset at hand (Vince D. Calhoun et al., 2009; Hyvärinen & Oja, 2000). Since here the main aim of the study was to investigate joint FC-SC task-dependent hybrid traits, we assessed the number of ICA components (ranging from 2 up to the dimension of the hybrid matrix after PCA reconstruction, see Fig. S1) that would maximize both the number of robust hybrid traits and task-based intra-class correlation (ICC) on their weights (see next section for details). This heuristic measure resulted in a optimal choice of 10 independent components (see Fig. S1).

**Task-based sensitivity**

We quantified whether a hybrid-trait was task-sensitive by using intra-class correlation (Bartko, 1966; Shrout & Fleiss, 1979). ICC is a widely used measure in statistics, normally to assess the percent of agreement between units (or ratings/scores) of different groups (or raters/judges) (McGraw & P, 1996). It describes how strongly units in the same group resemble each other. The stronger the agreement, the higher its ICC value. We used ICC to quantify to which extent the individual values of the weights of an hybrid trait could separate between subjects performing different tasks. Following this rationale, the different tasks are "raters" and "scores" given by the individual hybrid weights of the subjects. In this case, the higher the ICC, the more separable the different tasks across subjects and consequently the more task-dependent (i.e., higher changes in the weights) in the correspondent hybrid traits.

**Structural connectome randomization**

To avoid the possibility that the hybrid patterns were only driven by the functional profiles, we run the hybrid connICA with randomized structural connectomes. The edges of each individual SC were swapped 50,000 times, following the randomization technique proposed in (Goñi, Corominas-Murtra, Solé, & Rodríguez-Caso, 2010). This randomization preserves the main topological properties of the strurctural connectomes, such as size, density and degree-sequence (and hence degree distribution, (Goñi et al., 2010)). The chosen number of swaps (50,000) represents the best trade-off for this data between minimum number of swaps and maximum gain in dissimilarity of the randomized connectomes with respect to the original SCs (see Fig. S2).

## Results

The dataset used for this study consisted of structural and functional data from the 100 unrelated subjects in the Q3 release of the HCP (D. C. Van Essen et al., 2012; David C. Van Essen et al., 2013). For each subject, we estimated: 8 functional connectivity matrices, 1 corresponding to resting-state (by averaging the REST1_LR and REST1_RL FCs), 7 corresponding to each of the 7 tasks (by averaging LR and RL corresponding FCs, see Methods); 1 structural connectome, corresponding to the HCP DWI acquisition S1 (see Methods). The multimodal parcellation used here, as proposed by Glasser et al. (Glasser et al., 2016) , includes 360 cortical brain regions. We added 14 subcortical regions, hence producing functional connectome matrices (square, symmetric) of 374 x 374 (see Methods for details).

From the test-retest pool of 100 unrelated subjects (total of 800 FC matrices and 100 SC matrices), a bootstrap technique was used to accurately estimate the task-dependent hybrid traits. That is, for each run of hybrid connICA, a random cross-sectional sample comprising the functional-structural connectomes pairs of 80 subjects (10 subjects per task and resting state) was considered. This was meant to avoid results driven by a small subset of the population and to minimize redundancy in the SCs due to including same subjects performing different tasks.

The hybrid connICA procedure can be summarized as follows (Fig. 1, see also (Amico et al., 2017)): first, the upper and lower triangular parts of each individual FC and SC were vectorized and added to a matrix where rows are the subjects and columns are their full connectivity pattern; second, the ICA algorithm was run (100 runs, number of IC=10, see Methods) to extract the main hybrid (joint FC-SC) traits associated to the whole population; third, the most robust (appearing at least 50% of the times with correlation higher than 0.5, see Methods) and task-dependent components (as measured by intraclass correlation on the weights per different task/resting session, see Methods) were selected.

The hybrid connICA procedure resulted in two main task-sensitive hybrid traits (Fig. 2). The frequency of the averaged hybrid traits across runs were 90% and 89% respectively. That is, the main functional-structural patterns, common to the whole population, which weights change depending on the task that is being performed (high values of task-based intra-class correlation: 0.65 and 0.70, see Fig. A1, A2). A third robust averaged hybrid trait (64% frequency across runs) was obtained through hybrid connICA (Fig. S4), which however was not task-sensitive (ICC=0.16). Interestingly, this trait encompasses the main resting state networks, and corresponding within-network structural connections (Fig. S4).

Among the task-dependent hybrid traits, the functional part of the first trait mainly captures the within-connectivity of dorsal and visual networks, as well as interconnections among dorsal attentional, visual and sub-cortical networks (Fig.2). The structural part mainly evidences the within-network connectivity between these aforementioned three networks. The functional part of the second trait mainly represents the connectivity between the visual, attentional (dorsal and ventral), default-mode (DMN), fronto-parietal (FP) and subcortical networks (Fig. 2). The structural part mainly captures the within network connectivity between those and the limbic system. It is worth to mention here one of the advantages of the hybrid connICA procedure: that is, the hybrid traits represent joint structural-functional profiles learned from the whole population at the same time (the subject weights corresponding to the FC or SC are the same).

In order to assess the generalization of these results with respect to the gray matter parcellation used, we ran the same analyses with a lower resolution parcellation, namely Destrieux atlas (Destrieux et al., 2010; Fischl et al., 2004), as available in FreeSurfer software) which includes 164 brain regions. The two most frequent averaged hybrid traits (92% and 97% respectively) are shown in Fig. S5. Both hybrid traits were task-sensitive (ICC being 0.60 and 0.58 respectively). When comparing the hybrid traits obtained from both parcellations (Fig. 2 and Fig. S5), we observed a high resemblance from a RSNs (within and between) perspective.

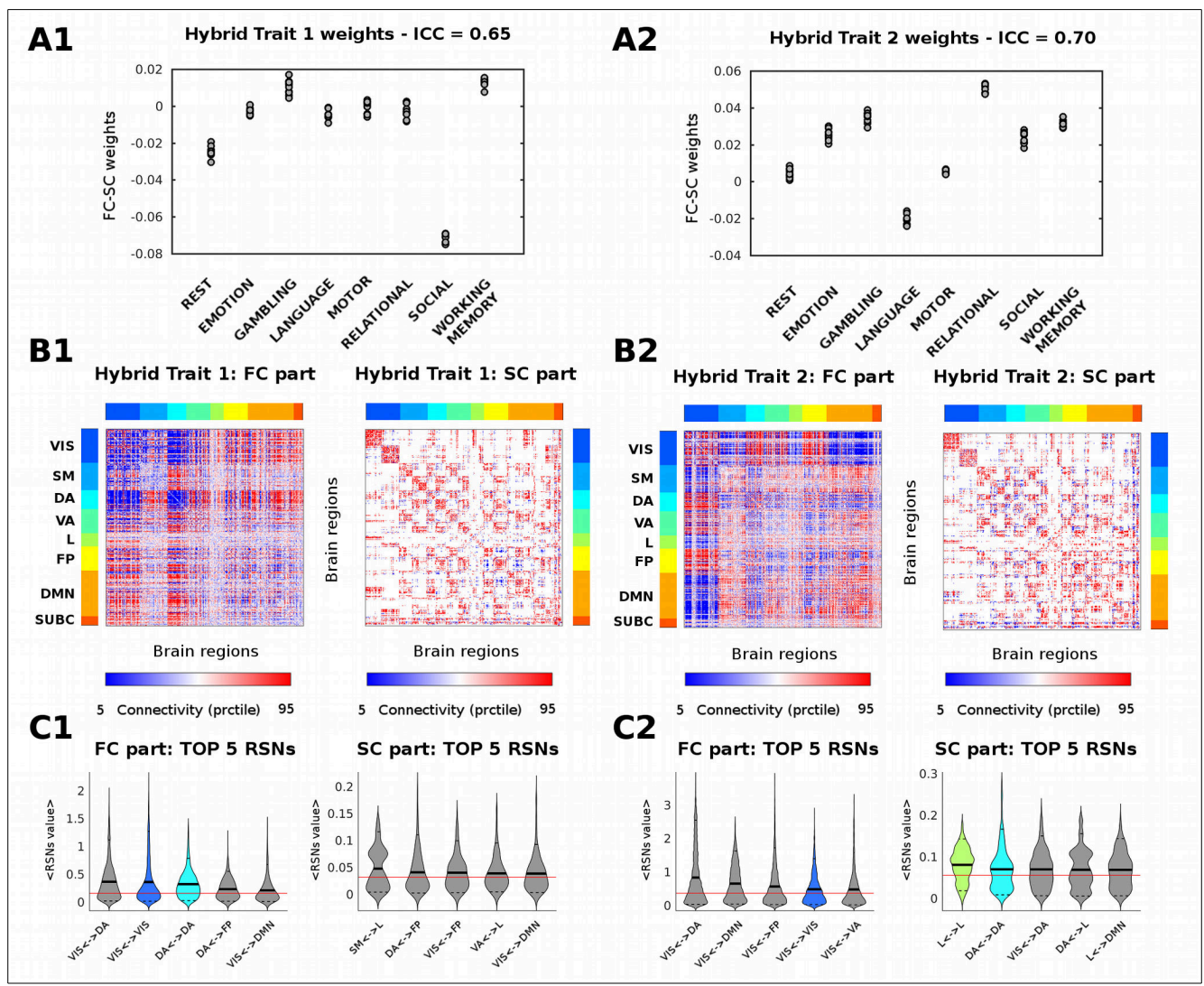

**Fig.2 Mapping of the main task-sensitive hybrid traits. A1-A2) Quantified presence of each hybrid trait on each individual functional connectome.** Subject weights are grouped according to each of the 7 tasks and resting state (10 subjects per task and resting state, see Methods). Task based intra-class correlation values are reported on top. **B1-B2) Visualization of the two hybrid traits associated to significant changes (as measured by ICC) between tasks and resting state.** For ease of visualization, the hybrid traits are split in two matrices, corresponding to the functional connectivity (FC) and structural connectivity (SC) patterns. The brain regions are ordered according to functional RSNs (Yeo et al., 2011): Visual (VIS), Somato-Motor (SM), Dorsal Attention (DA), Ventral Attention (VA), Limbic system

(L), Fronto-Parietal (FP), Default Mode Network (DMN), and for completeness, also subcortical regions (SUBC). **C1-C2) Violin plot of hybrid traits values for the top 5 RSNs.** The top 5 edge distributions per within or between RSNs interaction are shown separately for the FC and SC profiles. Each color indicates a different within RSN (as in B-C RSNs colorbar), while gray indicates edge values between RSN networks. The solid black lines of the violins depict the mean value of the distribution; the dashed black lines the 5 and 95 percentiles; the solid red line indicates the whole-brain mean value.

We then mapped each resulting hybrid connectivity profile onto a brain cortical surface (Fig.3). First we created a "hybrid mask" by taking only the most extreme edges in the FC and SC parts of the two hybrid traits (outside the $5^{th}$ and $95^{th}$ percentile of each distribution of values, see Fig. 3). That binary mask was then mapped onto a brain cortical mesh, to visualize the main hybrid circuitry involved in task-switching (Fig. 3). This allows to examine simultaneously functional nodes and structural pathways that are sensitive (i.e. differently engaged) along the tasks. In the case of the first trait, the dorsal cortical regions are more prominent, as expected, and their inter-hemispheric structural connections, as well as the fibers projecting from these regions to sub-cortices and frontal areas (Fig. 3). For the second trait, visual cortices are the most prominent functionally, as well as the pathways connecting DMN and FP regions (Fig. 3). Notably, none of these task-switching circuits (i.e. the joint masks in Fig.3) were found when robust hybrid traits were obtained from the randomization of the SCs (see Methods for details and Fig. S2). Indeed, it is noteworthy that the number of hybrid edges found in the joint FC-SC masks were significantly lower after randomization (see Fig. S3).

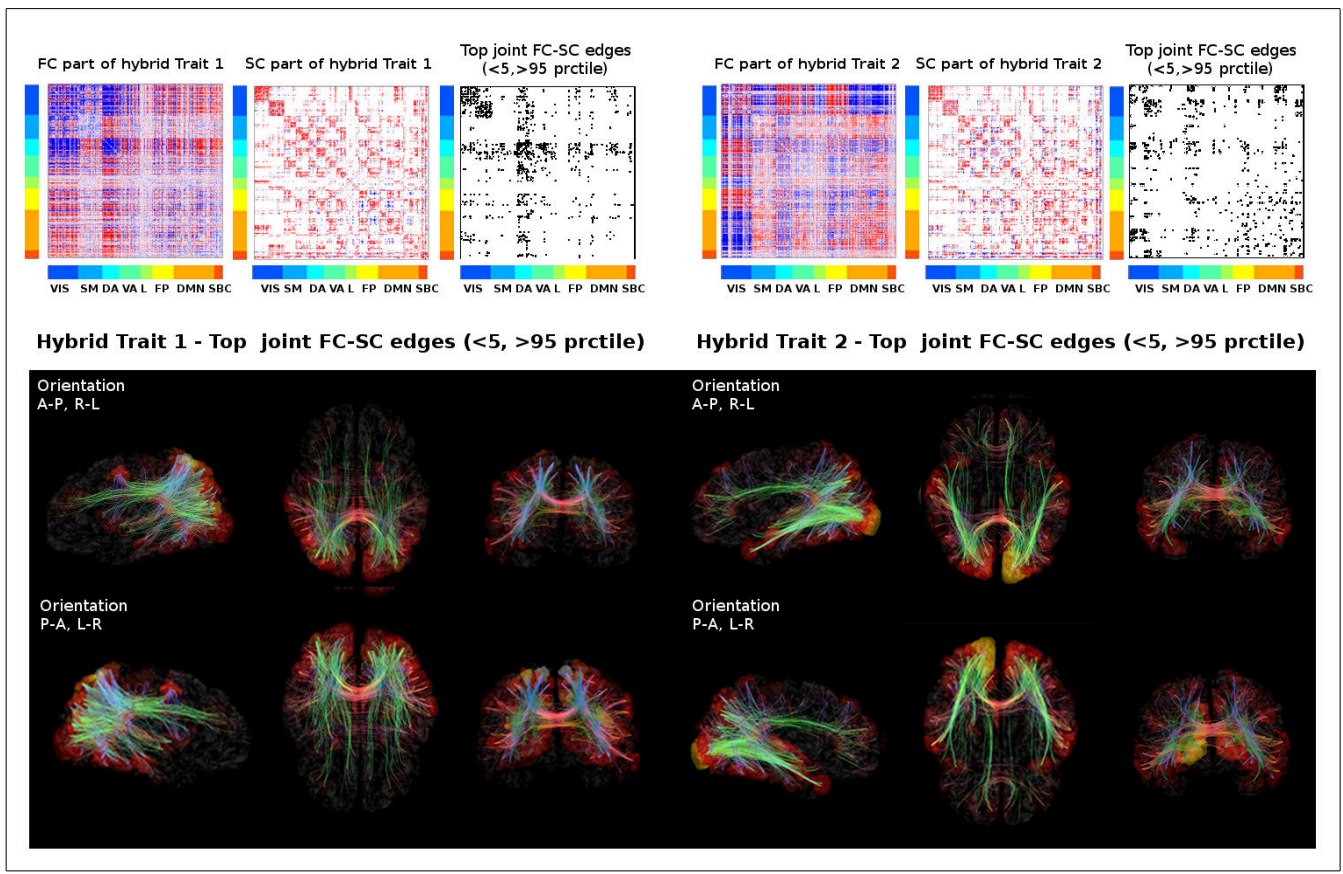

**Fig. 3. Visualizing task-sensitive joint functional-structural circuits in the human brain.** Top: The hybrid traits (split in functional and structural patterns) shown in Fig. 2, and the joint mask obtained from the product of the more extreme values (outside the [5,95] percentile range) in the corresponding FC and SC profiles. Bottom: The joint masks are projected onto brain renders, where tracts (color-coded by direction; Red:left-right; green: anterior- posterior; blue:superior-inferior.) represent non-zero edges in the masks, and nodal strength (sum over columns of the masks) is mapped onto the cortical meshes, from low strength (black) to high nodal strength (bright yellow). The brain renders were obtained with MRtrix3 (Tournier et al., 2012)

The analysis on the nodal strength on the joint mask allows for an assessment of the overall centrality of each region on the hybrid task-sensitive traits (Fig. 4). In the first hybrid trait, the main areas involved are the dorsal and ventral lobes, associated to attentional network connectivity, as expected. On the other hand, the left and right visual cortices dominate the nodal strength overview of the second hybrid trait (Fig. 4).

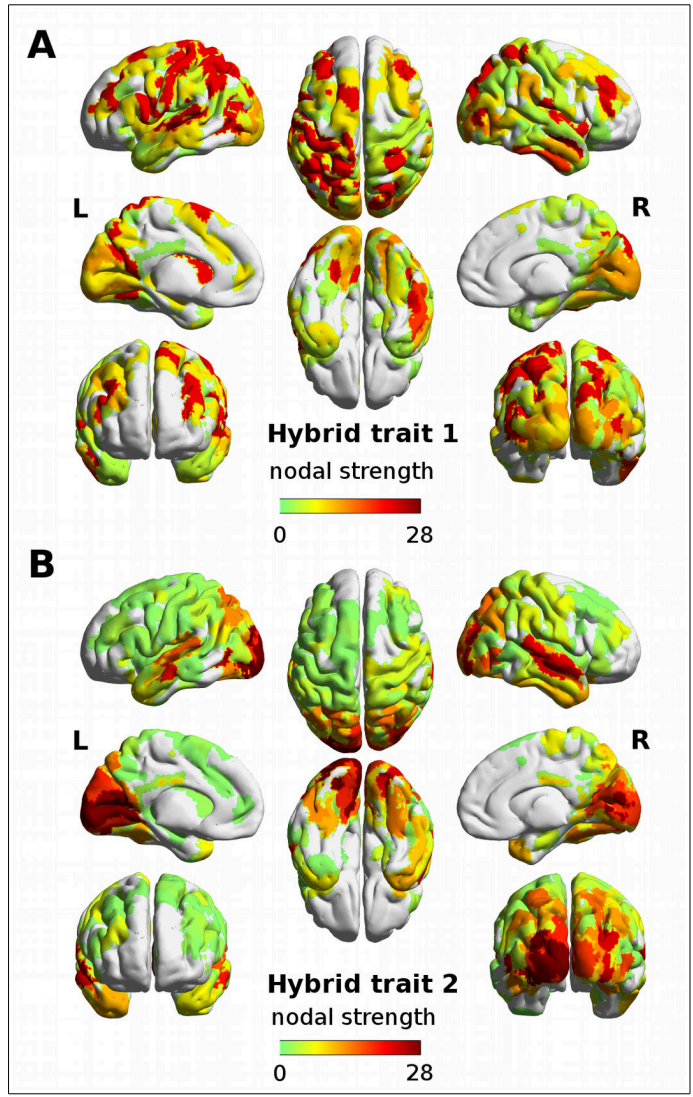

**Fig. 4. Node centrality of the hybrid task-sensitive traits.** The strength per region computed as sum of non-zero values in the joint FC-SC mask, for the A) first and B) second hybrid trait. The mask was obtained by taking only the most extreme edges in the FC and SC parts of the two hybrid traits (outside the $5^{th}$ and $95^{th}$ percentile of each distribution of values).

Note how the first trait mainly involves attentional-related areas in the dorsal and ventral lobes lobe, whereas both left and right visual cortices dominate in the second trait. The brain renders were obtained with BrainNet Viewer ("BrainNet Viewer: A Network Visualization Tool for Human Brain Connectomics,").

## Discussion

The investigation of the interaction between structural and functional connectivity layers in large-scale human brain networks is one of the current challenges in brain connectomics (Falcon et al., 2016; Fornito et al., 2016; Christopher J. Honey et al., 2010; Mišić et al., 2016). The difficulty of addressing this problem is manifold: from the different data processing to the huge amount of information of difficult interpretation, to the necessity of estimating individual weights of group-level structural-functional circuits.

Here we proposed a framework, named hybrid connICA (which expands on connICA (Amico et al., 2017)), that allows to extract, in a data-driven fashion, the most representative joint functional-structural (i.e. hybrid) patterns from a set of individual joint connectivity profiles (see Fig. 1). We tested this methodology on the HCP data benchmark, to retrieve the hybrid connectivity sub-system related to changes in functional tasks and resting state (Fig. 2, Fig. 3).

The hybrid connICA extracted two main task-dependent traits. The first, encompassing the within and between network connections of dorsal attentional and visual areas, as well as subcortical structures (Fig. 2). The second hybrid trait associated to task switching mainly specializes in the connectivity between the visual and frontoparietal, DMN and subcortical networks (Fig. 2).These findings are in line with previous research showing that these are the main areas (attentional cortices, DMN, visual and subcortical regions) undergoing major changes when passing from rest to task sessions (Cole, Bassett, Power, Braver, & Petersen, 2014; Fox & Raichle, 2007; Hasson, Nusbaum, & Small, 2009; Hermundstad et al., 2013; Tavor et al., 2016). Recent studies also reported the existence of an "intrinsic functional architecture" (Cole et al., 2014) that shapes brain's functional network architecture during task performance (Tavor et al., 2016). These findings suggest that a set of small but consistent changes in functional connectivity across tasks might distinguish task states from rest (Cole et al., 2014), and it might also allow to predict task states from the intrinsic resting-state organization (Tavor et al., 2016). We also assessed the reproducibility of the results by evaluating the two most frequent hybrid traits at a coarser-grain resolution (Fig. S5). Interestingly, a high resemblance from a RSNs (within and between) perspective was found, suggesting that the method presented here shows consistency in the results across brain spatial scales and resolutions.

Here, we took one step forward in this direction, by using hybrid connICA approach to map the main joint FC-SC circuitry involved in task switching (Fig.3). Interestingly, the intrinsic resting-state organization of a human brain both at the functional and structural level was also recovered, even though was not associated to changes across tasks. Indeed, one robust trait captured all the main functional and structural connections of RSNs blocks (hybrid trait 3, Fig. S4).

One benefit of this methodology resides in the possibility to extract and visualize "cities" (cortical functional nodes) and highways (structural connections) corresponding to specific subsystems simultaneously (Fig. 3). In the case of this work, functional-structural patterns that change depending on whether the subject is at rest or performing a specific task (Fig. 3). Notably, both hybrid traits capture two main aspects of brain network connectivity: integration (in the sense of functional interaction between networks) and segregation (expressed as main within network connectivity between structural circuits, Fig. 2 and Fig. S4) (Tononi, Sporns, & Edelman, 1994). It is worth to stress here that the resulting hybrid pattern are by all means entangled together. Notably, it is not likely to retrieve the same joint subsystems without the real connectivity structure (Fig. S3). When the SCs were randomized (see Fig. S3 and Methods), it was not possible to retrieve any of the hybrid joint pathways presented in Fig. 3 (i.e. the number of nonzero values in the FC-SC mask significantly lowered after randomization, see Fig. S3).

There are several advantages in applying a data-driven procedure such as hybrid connICA. The compression of the meaningful information into a few hybrid connectivity layers that are robust, independent and task-sensitive, is one of the major points. In addition, the subject weights associated to each hybrid trait are unique, meaning that there are single individual weights that allow to recover the structural-functional subsystem at the single-subject level (e.g. the ICA procedure was performed at once by concatenating structural and functional profiles, hence providing unique sets of weights associated to the FC and SC connectomes). This might ease inferences at the individual level with cognitive, genetic variables directly on the weights, and avoid multiple comparisons when working with multidimensional matrices. Our approach based on independent group-level hybrid traits with associated individual weights adds to recently proposed data-driven methods, where group-level orthogonal co-varying structural-functional patterns are extracted based on singular value decomposition (Mišić et al., 2016) While these approaches are focused on the integration of different modalities, frameworks such as canonical correlation analysis (Irimia & Van Horn, 2013) allow to disentangle which modalities are responsible for associations between different brain regions.

Here we showed an application of the hybrid connICA in disentangling task-dependent joint FC-SC circuits in healthy young adults. Next steps on using this framework will involve the investigation of hybrid patterns in clinical populations where heterogeneous individual structural damage is usually associated to a rich repertoire of different functional responses (such as in Parkinson, Alzheimer's disease, traumatic brain injury, disorders of consciousness, etc). This method can then provide a data-driven way to disentangle the main circuits associated to the disease (similarly to the functional connICA (Amico et al., 2017; Contreras et al., 2017)), while assessing structural and functional changes at the same time. This might also allow researchers to investigate and make inferences on the structural and functional circuitries involved, by compressing them into few hybrid traits. The flexibility of the method enables for extending the approach on layers other than the structural ones. For instance, one may substitute the SC layer proposed in this work with other network measures such as modularity, efficiency or search information. Also, one may assess the multimodal integration of different functional modalities such as electroencephalography at different frequency bands, or magnetoencephalography, among others.

This study has several limitations. The optimal size of the cohort for the extraction of the hybrid connICA components needs to be further investigated. Similarly, the best choice of the starting number of ICA components (here set to 10, see Fig. S1) and the threshold for the final selection of the most frequent components over multiple ICA runs (here set to 50%) depends on the research question at hand. It is important to have a priory hypotheses for filtering the robust traits to analyses. For instance, here we used intra-class correlation among tasks as a criterion. Also, in recent clinical studies, multi-linear models were used to associate the connectivity traits to crucial behavioral and/or clinical variables (Amico et al., 2017; Contreras et al., 2017).  Despite a state-of-the-art tractography (SIFT2, MRtrix3) algorithm was used in this study, further exploration of the sensitivity and specificity of the hybrid traits to different tractography could be performed. Finally, for the ICA extraction to work properly, we also strongly recommend for the range of the two connectivity profiles concatenated to be consistent across edges. Here we proposed the use of SC-based correlations. However, different normalizations could be applied for making SC and FC magnitudes comparable or at least more homogeneous (e.g. using absolute values, dividing by the maximum value,  or by applying L1 and L2 norms (V. D. Calhoun et al., 2006; Kessler et al., 2014)).

In conclusion, we here proposed a novel data-driven approach, hybrid connICA (successor of connICA (Amico et al., 2017)), to disentangle the most influential functional-structural connectivity patterns related to changes in brain networks across tasks and resting state. Our results shed light on the key hybrid circuitry (both functional and structural) involved in the differentiation of connectivity profiles across different tasks. By simultaneously extracting structural-functional subsystems, the proposed methodology might improve our understanding of connectivity changes associated to brain pathologies.

## Acknowledgments

Data were provided [in part] by the Human Connectome Project, WU-Minn Consortium (Principal Investigators: David Van Essen and Kamil Ugurbil; 1U54MH091657) funded by the 16 NIH Institutes and Centers that support the NIH Blueprint for Neuroscience Research; and by the McDonnell Center for Systems Neuroscience at Washington University. This work was partially supported by NIH R01EB022574 and by NIH R01MH108467 and by the Indiana Clinical and Translational Sciences Institute (Grant Number UL1TR001108) from the National Institutes of Health, National Center for Advancing Translational Sciences, Clinical and Translational Sciences Award.

## Code availability

The code used for extract hybrid connICA traits from joint FC-SC data is available on the CONNplexity lab website (*https://engineering.purdue.edu/ConnplexityLab*).

# Supplementary Materials

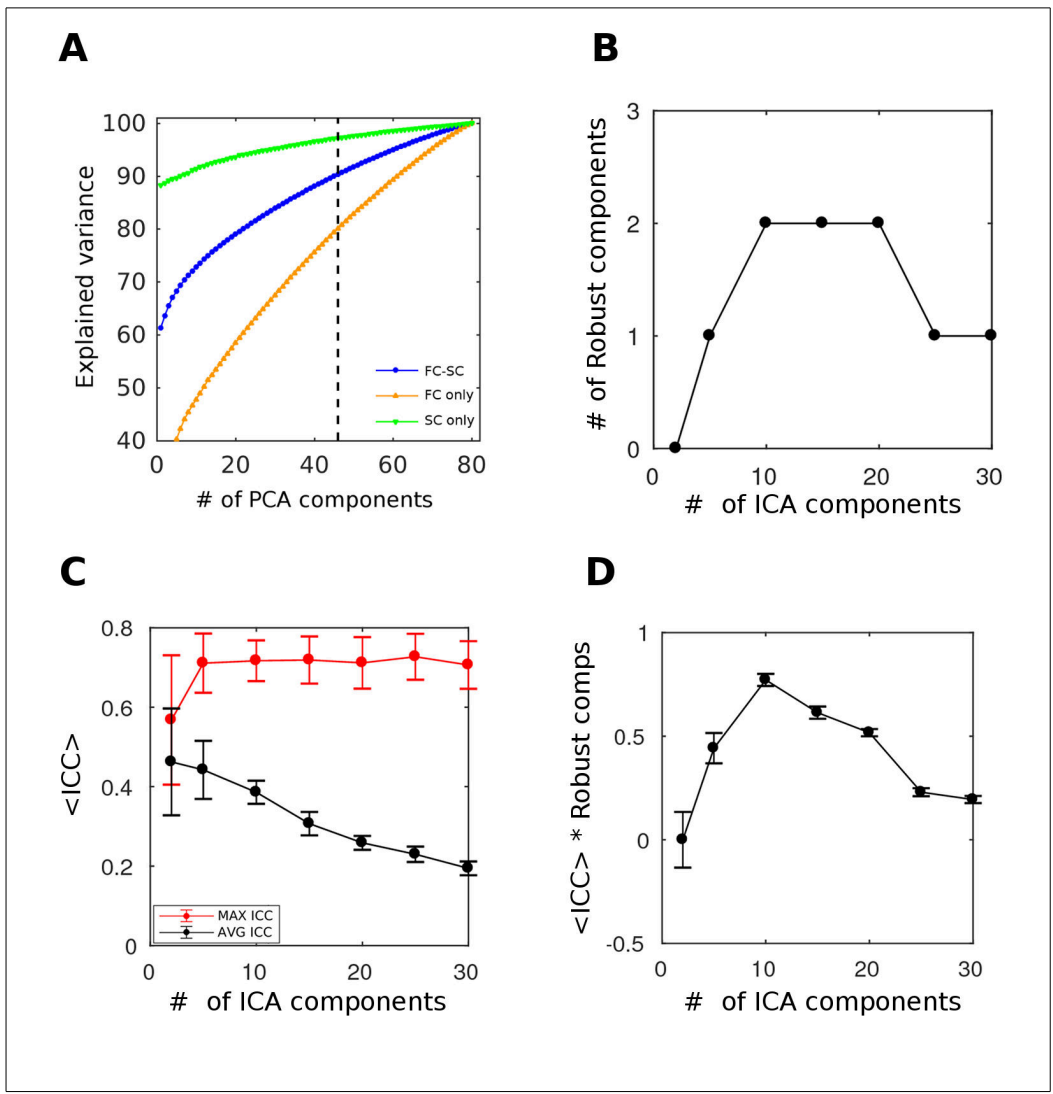

**Fig. S1. Robustness evaluation of hybrid connICA. A)** The cumulative explained variance for different components (from 1 to 80, being 80 the number of subjects (bootstrap, 100 runs, see Methods for details). Blue curve denote the cumulative explained variance for the joint FC-SC matrix, disentangled into FC (orange curve) and SC (green curve). The standard deviation across runs (not shown) was always lower than 1.5%. The dashed vertical black line represents the PCA cutoff at 90% of the variance on the joint scree plot (blue curve), at 45 components. Note that this cutoff on the joint FC-SC hybrid matrix captures about 95% of the variance on SC and about 80% of the variance of FC. **B)** ICA exploration was performed on the PCA reconstructed dataset to check the range with the highest number of robust hybrid traits (see Methods). **C)** Also, ICA exploration was performed in order to check the IC range that would maximize task intra-class correlation on the weights of the hybrid traits (see Methods). **D)** The function chosen to pick the optimal number of independent component was defined as the product between number of robust components and avg ICC value in the ICA set. In this study we set the optimal number of IC is to 10 accordingly.

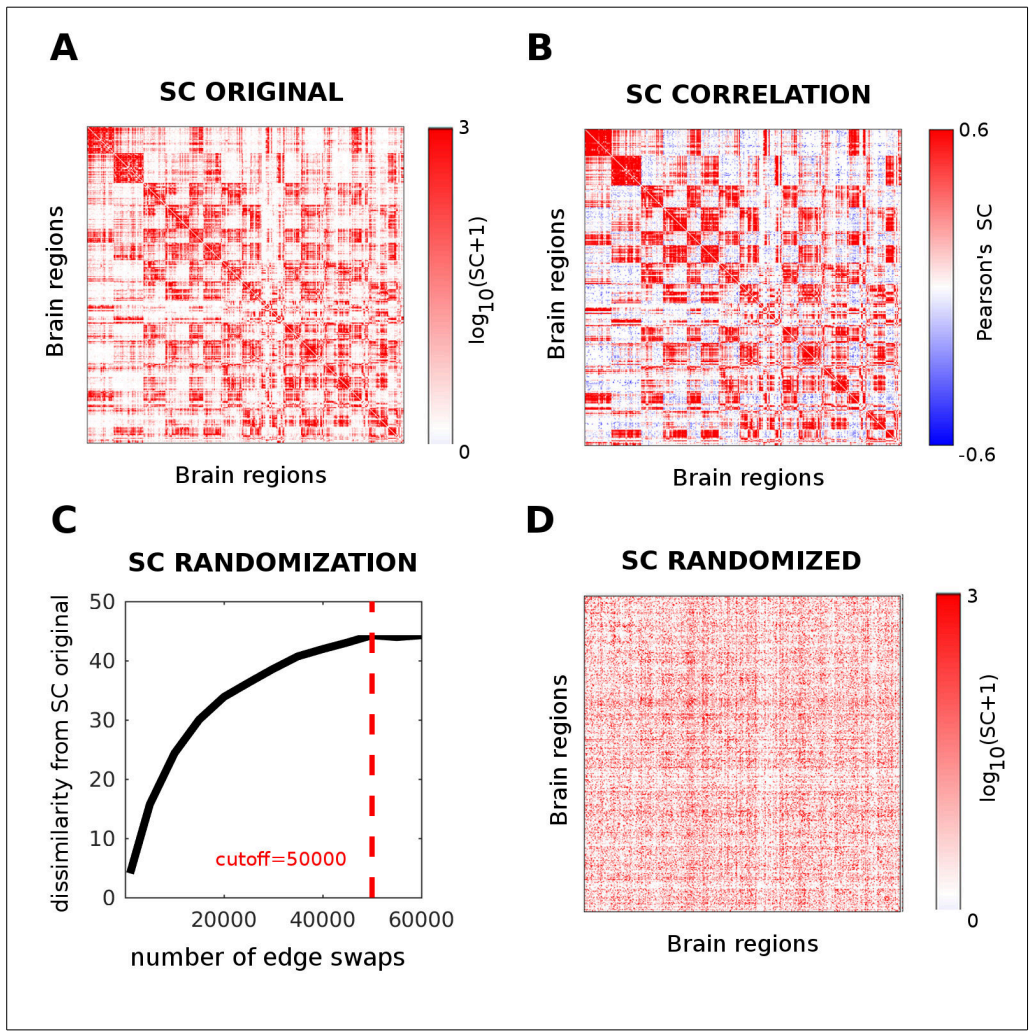

**Fig. S2. Illustration of the different structural connectivity configurations employed**. We started from one original individual structural connectome, SC (A); we then used for the hybrid connICA the spatial correlation matrix obtained from the original SC (B); we tested results of the procedure against driving forces related to FC by randomizing the original SC through edge swapping (C); the number of changes was fixed to 25000 swaps, since it provides a trade-off between minimum number of swaps and maximum gain in dissimilarity with the original SC (see Methods). Finally, the spatial correlation of the randomized SC patterns (D) was computed and inputed in the hybrid connICA.

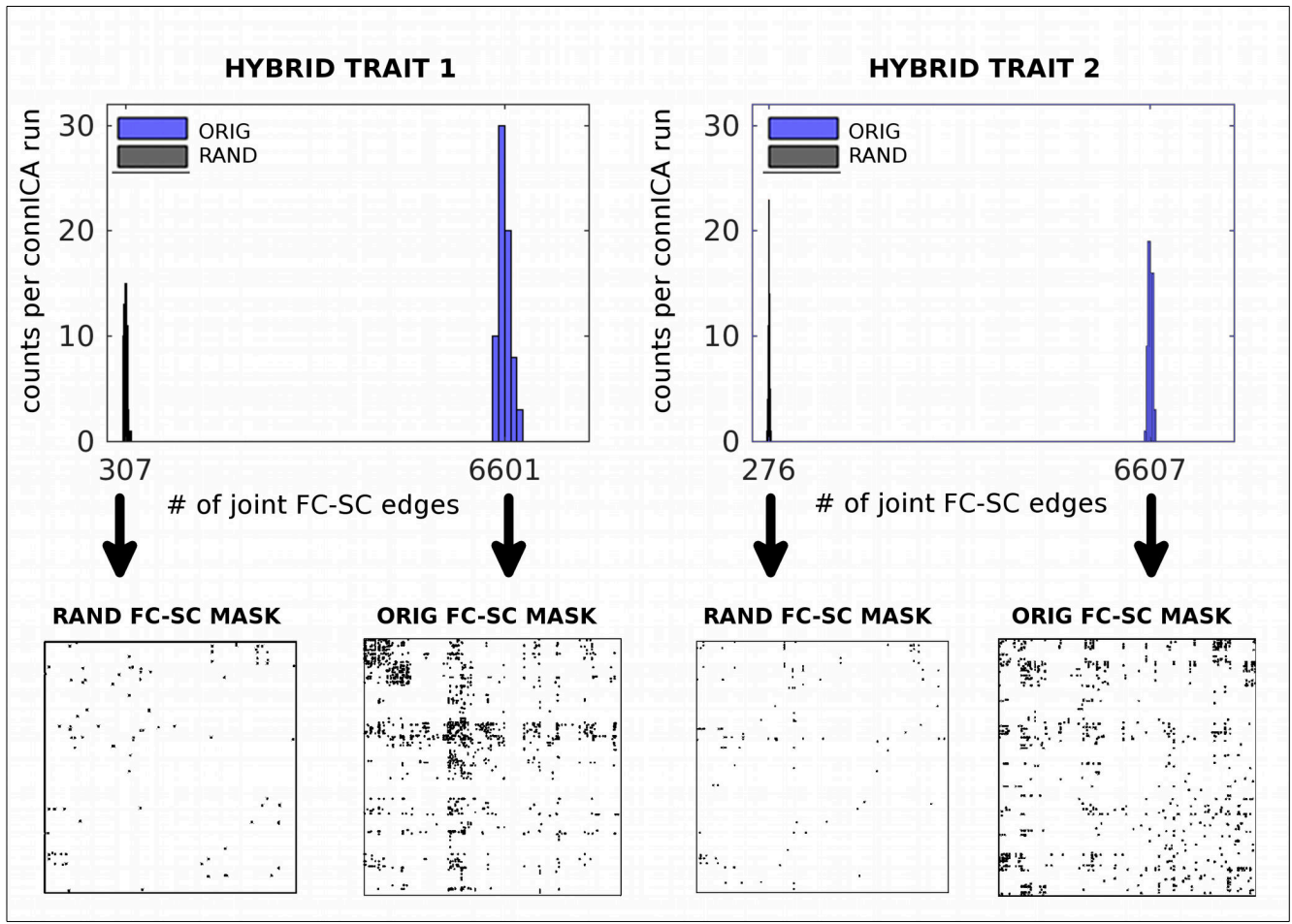

**Fig. S3. Comparisons between original hybrid traits and SC-randomized ones**. Top: For each of the 100 connICA runs, we counted the number of joint FC-SC edges in the mask (see also Fig.3), for the two original robust traits (blue histograms) and the two obtained after randomization of the structural connectomes (black histograms). Bottom: the joint masks obtained from the product of the more extreme values (outside the [5,95] percentile range) for the traits with randomized SC, and the original ones. Note how the number on nonzero elements significantly drops, hence making impossible the recovery of the hybrid circuitry depicted in Fig. 3 without the real structural connectivity profiles.

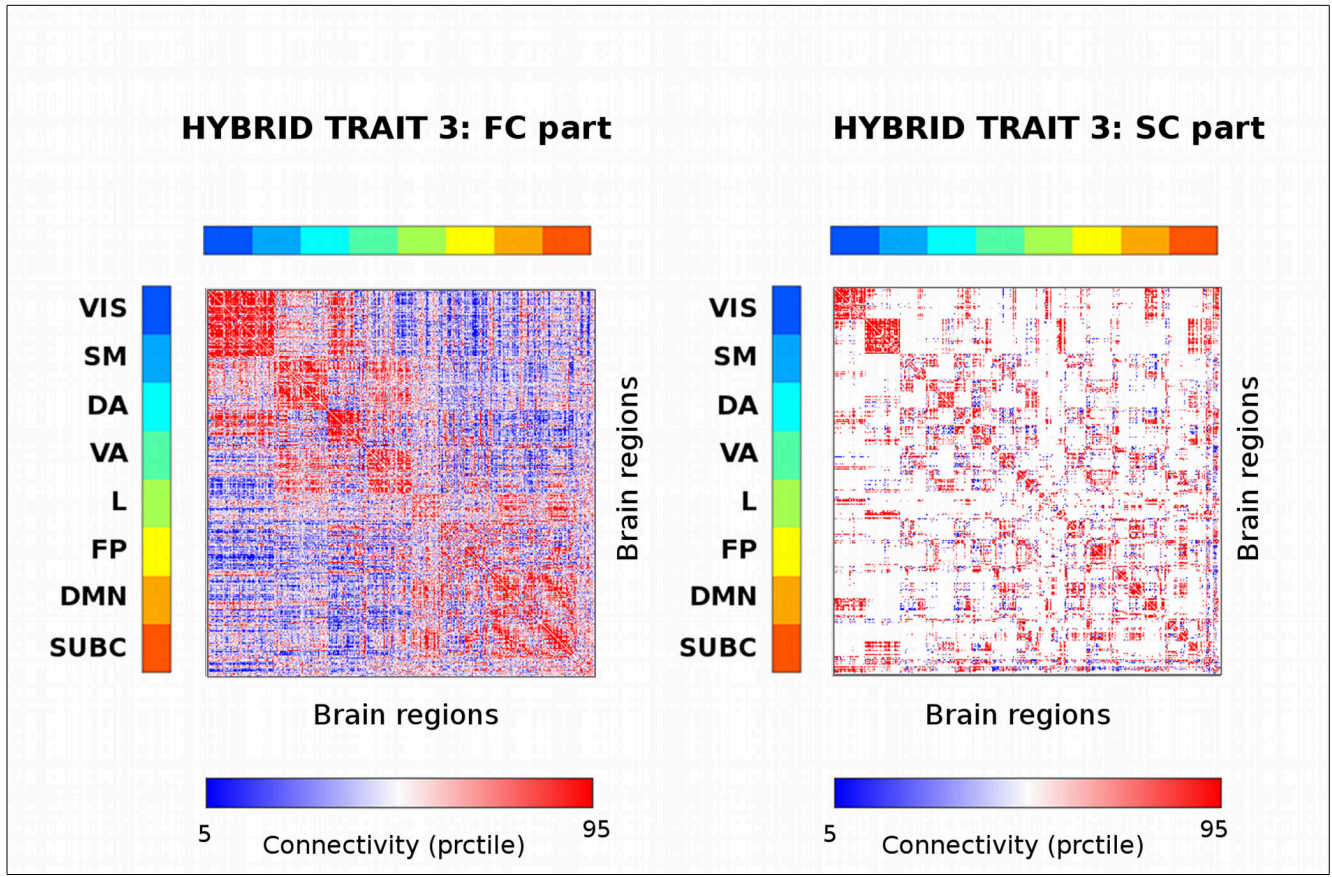

**Fig. S4. The third (non task-dependent) robust hybrid trait found by hybrid connICA** . The third robust trait extracted by hybrid connICA. Even if not related to task switching, it is representative of the intrinsic functional structural architecture of a human brain, since it mainly encompasses all the resting state networks functional blocks (left side of the plot), as well as the corresponding within network structural connections (right side of the plot).

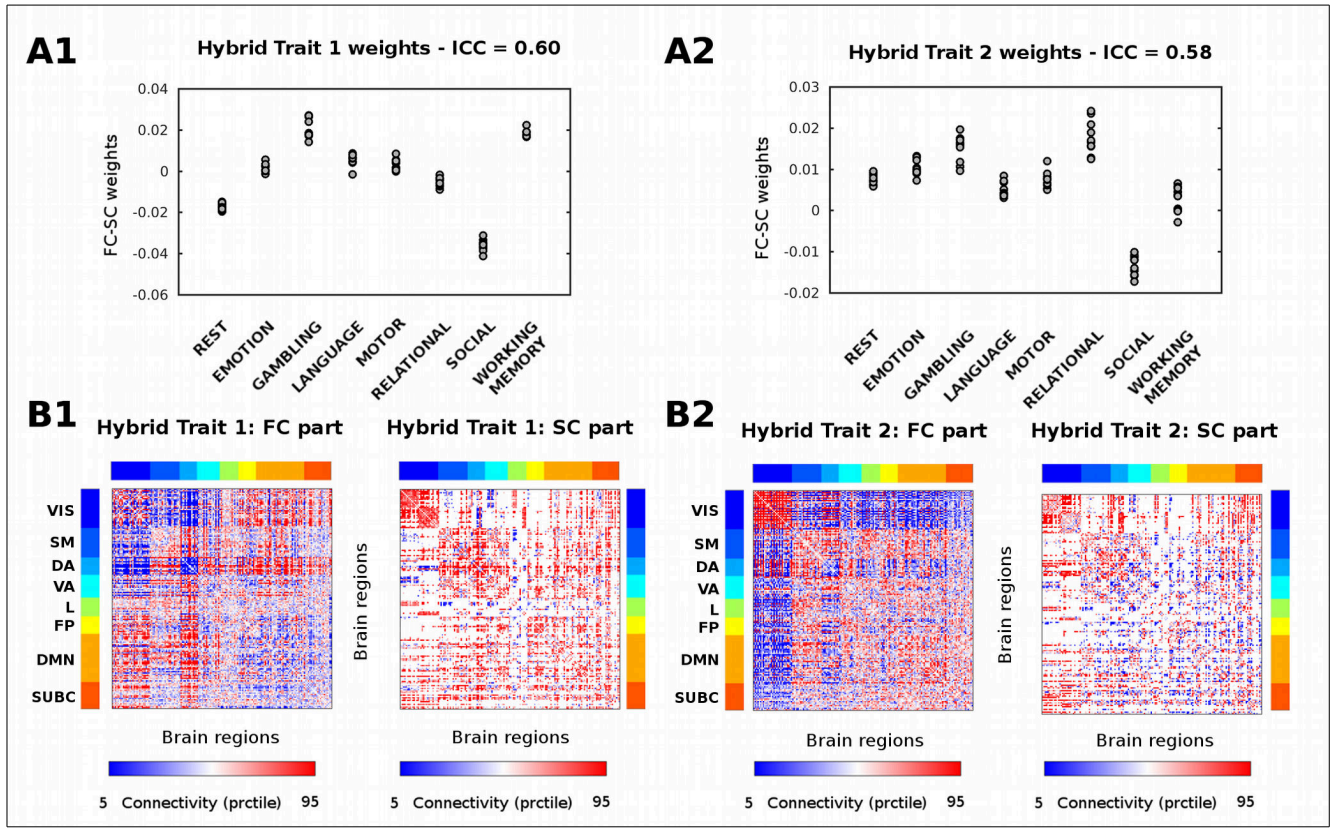

**Fig. S5 Mapping of the main task-sensitive hybrid traits (Destrieux atlas, 164 brain regions). A1-A2) Quantified presence of each hybrid trait on each individual functional connectome.** Subject weights are grouped according to each of the 7 tasks and resting state (10 subjects per task and resting state, see Methods). Task based intra-class correlation values are reported on top. **B1-B2) Visualization of the two averaged hybrid traits associated to significant changes (as measured by ICC) between tasks and resting state.** For ease of visualization, the hybrid traits are split in two matrices, corresponding to the functional connectivity (FC) and structural connectivity (SC) patterns. The brain regions are ordered according to functional RSNs: Visual (VIS), Somato-Motor (SM), Dorsal Attention (DA), Ventral Attention (VA), Limbic system (L), Fronto-Parietal (FP), Default Mode Network (DMN), and for completeness, also subcortical regions (SUBC). Note that very similar traits were found when using a fine-grained parcellation (see Fig. 2, main manuscript).